\documentclass[letter, twocolumn]{aastex631}
\usepackage[utf8]{inputenc}
\usepackage[english]{babel}
\usepackage[T1]{fontenc}

\usepackage{color}
\usepackage{url}
\usepackage{natbib}
\usepackage{comment}
\shorttitle{X-ray polarization in Mrk~421}
\shortauthors{Di Gesu et al.}
\usepackage{soul}

\newcommand{\virg}[1]{``#1"}
\def \ergsc{\hbox{erg s$^{-1}$ cm$^{-2}$}}

\def \ergsch{\hbox{erg s$^{-1}$ cm$^{-2}$ Hz$^{-1}$}}

\def \nh {\hbox{ $N{\rm _H}$ }}
\def \colc {cm$^{-2}$}

\def \fwhm {{\em FWHM}}

\def \chis{$\chi ^{2}$}
\def \chidof{$\frac{\chi ^{2}}{\rm d.o.f}$}

\newcommand{\I }{{\it I}}
\newcommand{\Q }{{\it Q}}
\newcommand{\U}{{\it U}}

\begin{document}


\title{The X-ray Polarization View of Mrk~421 in an Average Flux State as Observed by the {\it Imaging X-ray Polarimetry Explorer}}
\author{Laura Di Gesu}
\correspondingauthor{Laura Di Gesu}
\email{laura.digesu@est.asi.it}
\affiliation{ASI -Agenzia Spaziale Italiana, Via del Politecnico snc, 00133 Roma, Italy}
\author{Immacolata Donnarumma}
\correspondingauthor{Immacolata Donnarumma}
\email{immacolata.donnarumma@asi.it}
\affiliation{ASI - Agenzia Spaziale Italiana, Via del Politecnico snc, 00133 Roma, Italy}
\author{Fabrizio Tavecchio}
\correspondingauthor{Fabrizio Tavecchio}
\email{fabrizio.tavecchio@inaf.it}
\affiliation{INAF Osservatorio Astronomico di Brera, Via E. Bianchi 46, 23807 Merate (LC), Italy}
%
%

\author{Iv\'{a}n Agudo}
\affiliation{Instituto de Astrof\'{i}sica de Andaluc\'{i}a-CSIC, Glorieta de la Astronom\'{i}a s/n, 18008, Granada, Spain}
\author{Thibault Barnounin}
\affiliation{Universit\'{e} de Strasbourg, CNRS, Observatoire Astronomique de Strasbourg, UMR 7550, 67000 Strasbourg, France}
\author{Nicol\`{o} Cibrario}
\affiliation{Dipartimento di Fisica, Universit\`{a} degli Studi di Torino, Via Pietro Giuria 1, 10125 Torino, Italy}
\affiliation{Istituto Nazionale di Fisica Nucleare, Sezione di Torino, Via Pietro Giuria 1, 10125 Torino, Italy}
\author{Niccol\`{o} Di Lalla}
\affiliation{Department of Physics and Kavli Institute for Particle Astrophysics and Cosmology, Stanford University, Stanford, California 94305, USA}
\author{Alessandro Di Marco}
\affiliation{INAF Istituto di Astrofisica e Planetologia Spaziali, Via del Fosso del Cavaliere 100, 00133 Roma, Italy}
\author{Juan Escudero}
\affiliation{Instituto de Astrof\'{i}sica de Andaluc\'{i}a-CSIC, Glorieta de la Astronom\'{i}a s/n, 18008, Granada, Spain}
\author{Manel Errando}
\affiliation{Physics Department and McDonnell Center for the Space Sciences, Washington University in St. Louis,
St. Louis, MO63130, USA}
% ...
%

\begin{abstract}
Particle acceleration mechanisms in supermassive black hole jets, such as shock acceleration, magnetic reconnection, and turbulence, are expected to have observable signatures in the multi-wavelength polarization properties of blazars. 
The recent launch of the {\it Imaging X-ray Polarimetry Explorer} ({\it IXPE}) enables us, for the first time, to use polarization in the X-ray band (2-8 keV) to probe the properties of the jet synchrotron emission in high-frequency-peaked BL Lac objects (HSPs).
We report the discovery of  X-ray linear polarization  (degree $\Pi_{\rm x}=15\pm$2\% and electric-vector position angle $\psi_{\rm x}=35\pm4\degr$) from the jet of the HSP Mrk~421 in an average X-ray flux state. At the same time, the degree of polarization at optical, infrared, and millimeter wavelengths was found to be lower by at least a factor of 3. 
During the {\it IXPE} pointing, the X-ray flux of the source increased by a factor of 2.2, while the polarization behavior was consistent with no variability.
The higher level of $\Pi_{\rm x}$ compared to longer wavelengths, and the absence of significant polarization variability, suggest a shock as the most likely X-ray emission site in the jet of Mrk 421 during the observation. The multiwavelength polarization properties are consistent with an energy-stratified electron population, where the particles emitting at longer wavelengths are located farther from the acceleration site, where they experience a more disordered magnetic field.

\end{abstract}

\newpage

\section{Introduction} 

Relativistic extragalactic jets expelled by active galactic nuclei (AGNs) are the most powerful persistent emitters in the Universe \citep[e.g.,][]{blandford19}. The understanding of their physical structure, dynamics, and impact on the surrounding environment plays a fundamental role in our view of black holes, galaxies, and clusters. The current framework (e.g., \citealt{tchekhovskoy2011, komissarov2007}) assumes that jets are accelerated and collimated by magnetic stresses acting close to the central black hole. As the jet propagates, magnetic flux is progressively converted into kinetic flux, accelerating the flow. The dissipation of a fraction of the (magnetic and/or kinetic) energy flux provides the energy required to accelerate particles to ultrarelativistic energies. These particles produce extremely luminous emission across the electromagnetic spectrum. 

Jets are most prominent when the outflow velocity is nearly aligned with the line-of-sight and the emission is consequently boosted by relativistic effects. This occurs in blazars, AGNs characterized by powerful, variable (with time-scales as short as a few minutes in the most extreme cases), non-thermal emission extending from radio waves to $\gamma$ rays (e.g. \citealt{2017SSRv..207....5R}). The spectral energy distribution (SED) of blazars displays two broad components that are ascribed to synchrotron radiation peaking at IR to X-ray frequencies and Compton scattering with maximum at $\gamma$-ray energies (but see, e.g., \citealt{hadronic} for an alternative hadronic interpretation), respectively. Blazars in which the synchrotron luminosity peaks at X-ray energies are termed high-synchrotron-peaked BL Lacertae objects (HSPs or HBLs), the prototype of which is Mrk~421.
\begin{table*}[ht]
\caption{Log of the observations of Mrk~421 used in this work.} 
\centering  
\begin{tabular}{lcccc}
\hline
Telescope & Band & Dates & Radio/Optical flux density & Radio/Optical Polarization\\
& (eV) & (YYYY-MM-DD) & (\ergsch{})  & $\Pi$(\%) \quad $\psi$ ($\degr$)\\
\hline
Nordic Optical Telescope & R: 1.9  & 2022-05-04 & $(1.9\pm0.03) \times 10^{-25}$ & $2.9\pm0.1$ \quad $21\pm 1$\\
IRAM & $3.5 \times 10^{-4}$ & 2022-05-05 &$(2.8 \pm 0.1) \times 10^{-24}$  & $3.0\pm0.7$ \quad $57\pm 7$\\
Perkins & H: 0.8  & 2022-05-05 &$(3.94 \pm 0.03) \times 10^{-25}$  & $0.9\pm0.3$ \quad $39\pm 10$\\
Perkins & H: 0.8 & 2022-05-06 & $(4.48 \pm 0.03) \times 10^{-25}$  & $2.1\pm0.6$ \quad $39\pm 10$\\
Perkins & H: 0.8 & 2022-05-07 & $(4.20 \pm 0.04) \times 10^{-25}$ & $2.2\pm0.5$ \quad $32\pm 8$\\
IRAM & $3.5 \times 10^{-4}$  & 2022-05-07 & $(3.8 \pm 0.2) \times 10^{-24}$ & $3.8\pm0.7$ \quad $52\pm5$\\
IRAM & $9.5 \times 10^{-4}$ & 2022-05-07 & $(2.5 \pm 0.1) \times 10^{-24}$ & $\leq 6$ \quad -\\
\hline
Telescope & Energy range & Dates & X-ray flux & X-ray Polarization\\
& (keV) & (YYYY-MM-DD) & (\ergsc)  & $\Pi_{\rm x}$(\%) \quad $\psi_{\rm x}$ ($\degr$)\\
\hline
{\it Swift-XRT} & 0.3-10.0 & 2022-05-03 & $(3.50 \pm 0.05) \times 10^{-10}$ & - 
\quad -\\
{\it XMM-Newton} & 0.5-10.0 & 2022-05-03 & $(3.04 \pm 0.01) \times 10^{-10}$ & - \quad - \\
 {\it IXPE} & 2.0-8.0 & 2022-05-04 & $(8.67 \pm 0.03) \times 10^{-11}$  & $15\pm2$ \quad $35\pm4$ \\
{\it NuSTAR} & 3.0-30.0 & 2022-05-04 & $(7.21 \pm 0.11) \times 10^{-10}$  & - \quad - \\
{\it Swift-XRT} & 0.3-10.0 & 2022-05-05 & $(4.16 \pm 0.08) \times 10^{-10}$ & - \quad - \\
{\it Swift-XRT} & 0.3-10.0  & 2022-05-07 & $(5.93 \pm 0.09) \times 10^{-10}$  & - \quad - \\
\hline
\end{tabular}
\label{obs.tab}
\end{table*}

%
The precise nature of the processes responsible for the acceleration of high-energy particles remains unclear (see \citealt{matthews2020} for a review). Shocks are likely to occur in blazar jets, where particles can be accelerated through the well-studied diffusive shock acceleration (DSA) process \citep[e.g.,][]{blandford87}. However, shocks are unlikely to be efficient if the jet is highly magnetized \citep{sironi15}. In that case, it is more likely that particles are accelerated through conversion of magnetic energy in reconnection events (e.g. \citealt{sironi14}). Polarimetric analysis of the synchrotron emission, which provides information on the geometry and structure of magnetic fields, can help to specify which acceleration process dominates \citep[e.g.,][and references therein]{tavecchio21}. Measurement of linear polarization at X-ray energies from HSPs is particularly important, since it probes the highest-energy electrons, which radiate close to the acceleration site. Simultaneous observations of polarization at optical and longer wavelengths then allow comparisons of the physical conditions experienced by these electrons with those of regions where only lower-energy particles are present. The results can then be used to test the proposed models for particle acceleration in blazar jets.\\
 Previous polarization measurements of blazars have been limited to frequencies below those where the synchrotron emission of an HSP peaks. The Imaging X-ray Polarimetry Explorer \citep{Weisskopf2022}, a joint  NASA and Italian Space Agency (ASI) mission launched on 2021 December 9, is the first  telescope capable of measuring the polarization of blazars at X-ray energies. The {\it IXPE} payload comprises three X-ray telescopes with three Gas Pixel Detector Units (DUs, \citealt{gpd, gpd2}) in the focal planes, sensitive to linear polarization in the 2-8 keV energy band. Liodakis et al.\ (2022, submitted) have reported detection of X-ray polarization by {\it IXPE} in the HSP Mrk~501.
Here we report the first {\it IXPE} observation of Mrk~421 (redshift $z=0.0308$).
This source is a relatively nearby HSP that has been intensively studied at many wavelengths \citep[e.g.,][]{abdo11,Acciari21}.
The source possesses significant variability in the optical parameters of polarization. For instance, \citet{marscher21} reported that during intensive optical polarization sampling in April-May 2017, Mrk~421 exhibited variations in polarization degree $\Pi_{\rm o}$ between 6 and 11\% and changes in
electric vector position angle $\Psi_{\rm o}$ by $\sim$50$\degr$ on a timescale of 1-2 days.
Mrk~421 is among the first blazars detected at both GeV (by EGRET onboard the {\it Compton} Gamma Ray Observatory; \citealt{lin1992}) and TeV energies (by the Whipple Observatory; \citealt{punch1992}). It is well-monitored in the X-ray band, where the synchrotron SED peaks at a high flux level, making it a prime target for linear polarization observations by {\it IXPE}.
%


\section{X-ray Polarization Analysis}
\label{pol}
\subsection{Time-averaged polarization properties}
{\it IXPE} observed Mrk~421 from 10:00 UTC on 2022 May 4 to 11:10 UTC on May 6, with a live-time exposure of 97 ks, alongside a multi-wavelength campaign. The latter included observations with the {\it Swift X-Ray Telescope} (XRT), {\it XMM-Newton} ({\it XMM}), and the {\it Nuclear Spectroscopic Telescope Array} ({\it NuSTAR}) at X-ray energies, the Nordic Optical Telescope at the optical R band, Boston University's Perkins Telescope at infrared (IR) wavelengths, and the Institut de Radioastronomie Millim\'etrique ({\it IRAM}) 30 m telescope at short millimeter wavelengths. The  optical, IR, and radio flux densities and linear polarization measurements are summarized in Table \ref{obs.tab}. The data reduction procedure for all of the telescopes is described in Appendix \ref{sec:IXPE_data}. At the epoch of the {\it IXPE} observation, the 0.3-10 keV X-ray flux of Mrk~421 was $\sim3.5\times 10^{-10}$ \ergsc, which is near its typical flux state \citep{liodakis2019}. As a comparison, during occasional flaring events reaching gamma-ray energies, the X-ray flux of Mrk~421 can be ten times higher \citep{donna2009}.

We have determined the 2-8 keV polarization of Mrk~421 using (1) the \citet{Kislat2015} method within the ixpeobssim simulation and analysis software \citep{baldini2022}, (2) a spectro-polarimetric fit of the {\it IXPE} data alone, (3) a spectro-polarimetric fit of combined quasi-simultaneous data from {\it IXPE}, the {\it XMM} EPIC-pn camera, and {\it NuSTAR}, (4) a maximum likelihood spectro-polarimetric (MLS) fit of the IXPE data, implemented within the multinest algorithm, and, (5) a maximum-likelihood calculation of the polarization properties based on the method of \citet{likelihood}. While the \citet{Kislat2015} approach and the maximum likelihood calculation are model independent, spectropolarimetric and MLS fits include modeling of the energy spectrum, allowing us to test whether the determination of the polarization properties is affected by the modeling of the spectral curvature.
For Mrk~421, we find a minimum detectable polarization at 99\% significance (MDP99) of 6.4\% in the 2-8 keV band. The measured polarization parameters from the three combined detector units are $\Pi_{\rm x}=15\pm2$ \% and $\psi_{\rm x}=+35\pm 4\degr$. 
Given the MDP99 level of the observation, the 2-8 keV polarization of Mrk~421 is detected at a confidence level of $>$99.99\% (i.e., at a significance of $7\sigma$). By computing the polarization properties in three energy bins, i.e., 2.0-3.0 keV, 3.0-4.0 keV, and 4.0-8.0 keV, we did not observe any significant variation of the polarization properties with photon energy within maximum error bars of $\pm5$\% and $\pm8\degr$ for the polarization degree and polarization angle, respectively.

\begin{figure}
\resizebox{\hsize}{!}{\includegraphics[scale=1]{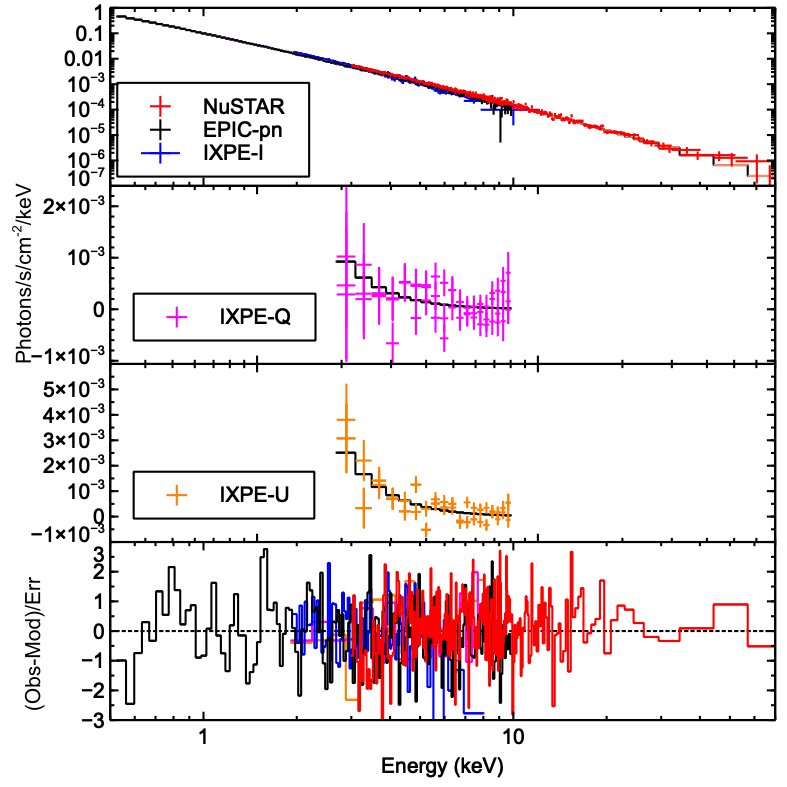} }
 \caption {Spectro-polarimetric fit of the {\it NuSTAR} (red), {\it XMM} EPIC-pn (black), and {\it IXPE} (blue, magenta, and orange) spectra. From the top to the bottom, we show the fit of the $I$, $Q$, $U$ spectra and the residuals of the best fit model (solid line). The spectral channels have been rebinned for clarity.}
\label{specpol.fig}
\end{figure}
%
%
%
We have also determined the degree $\Pi_{\rm x}$ and position angle $\psi_{\rm x}$ of polarization via a spectro-polarimetric fit in Xspec \citep{xspec} which implements the approach of
\citet{Strohmayer2017}.
In all of the fits, we included Galactic photo-electric absorption along the line of sight toward Mrk~421 (\nh=$1.34 \times 10^{20}$ \colc; \citealt{nh}) using the TBABS model, with Fe abundances set according to
\citet{abund_wilm}. As a first test, we considered only the {\it IXPE} data (i.e., 3 Stokes parameters for 3 DUs). In the fit we considered multiplicative constants, free to vary, to account for inter-calibration differences among the three DUs. The shape of the $I$ spectra is best fit (\chidof=293/267) by a power law with photon index $\Gamma=2.97\pm0.01$. By multiplying the model for the $I$ spectra by a constant polarization model (polconst model in Xspec), we obtained a statistically acceptable spectropolarimetric fit of the $I$, $Q$, and $U$ spectra (\chidof=841/795). The fluxes measured by the three IXPE DUs agree with each other within 12\%, with the differences explained by uncertainties in the {\it IXPE} effective area at the time of observation. The estimated polarization degree and position angle,
$\Pi_{\rm x}=16\pm2\%$ and $\psi_{\rm x}=34\pm3\degr$, are consistent with the ixpeobssim result.

Next, we used the simultaneous {\it XMM-Newton} and {\it NuSTAR} data to refine the constraints on the broad-band spectral shape and check whether the derived polarization is sensitive to the addition of broad-band spectral curvature in the model of the energy spectrum.
We initially fit the data from {\it XMM} EPIC-pn, {\it NuSTAR} FPMA, and {\it NuSTAR} FPMB plus the three {\it IXPE} Stokes $I$ spectra with a power-law model subject to photo-electric absorption (see above). We fixed the normalization and photon index $\Gamma$ for all of the instruments, while we allowed five multiplicative constants to vary to account for cross-calibration between the instruments and source variability between the observations (see Table \ref{obs.tab}). This model does not provide a statistically acceptable fit to the broadband data (\chidof=2764/611), which indicates that, when considering
an energy range broader than that of {\it IXPE}, a more complex model of the spectral curvature is needed. The broad-band spectrum of Mrk~421 is better fit by a log-parabolic model \citep{logpar}, in which the photon index varies as a log-parabola in energy i.e.
$$
N(E)=K(E/E_{\rm p})^{(\alpha-\beta \log{(E/E_{p}}))},
$$
where
the pivot energy $E_{\rm p}$ is a scaling factor, $\alpha$ describes the spectral slope at the pivot energy, $\beta$ describes the spectral curvature and $K$ is a normalization constant. This spectral shape is typical for HSP blazars, including Mrk~421, both in quiescence and in flaring states \citep[e.g.,][]{balokovic2016,donna2009}. As the photon index varies with the energy in this model, the choice of the reference energy $E_{\rm p}$ changes the determination of $\alpha$. In our fits, we set the pivot energy $E_{\rm p}$ to 5.0 keV (e.g., \citealt{balokovic2016}), while the $\alpha$ and $\beta$ parameters were allowed to vary, although they were coupled across the three telescopes.  With our setup, $\alpha$ approximates the photon index in the 3.0-7.0 keV range (i.e., the core of the IXPE band). %

\begin{figure}
\resizebox{\hsize}{!}{\includegraphics[scale=1]{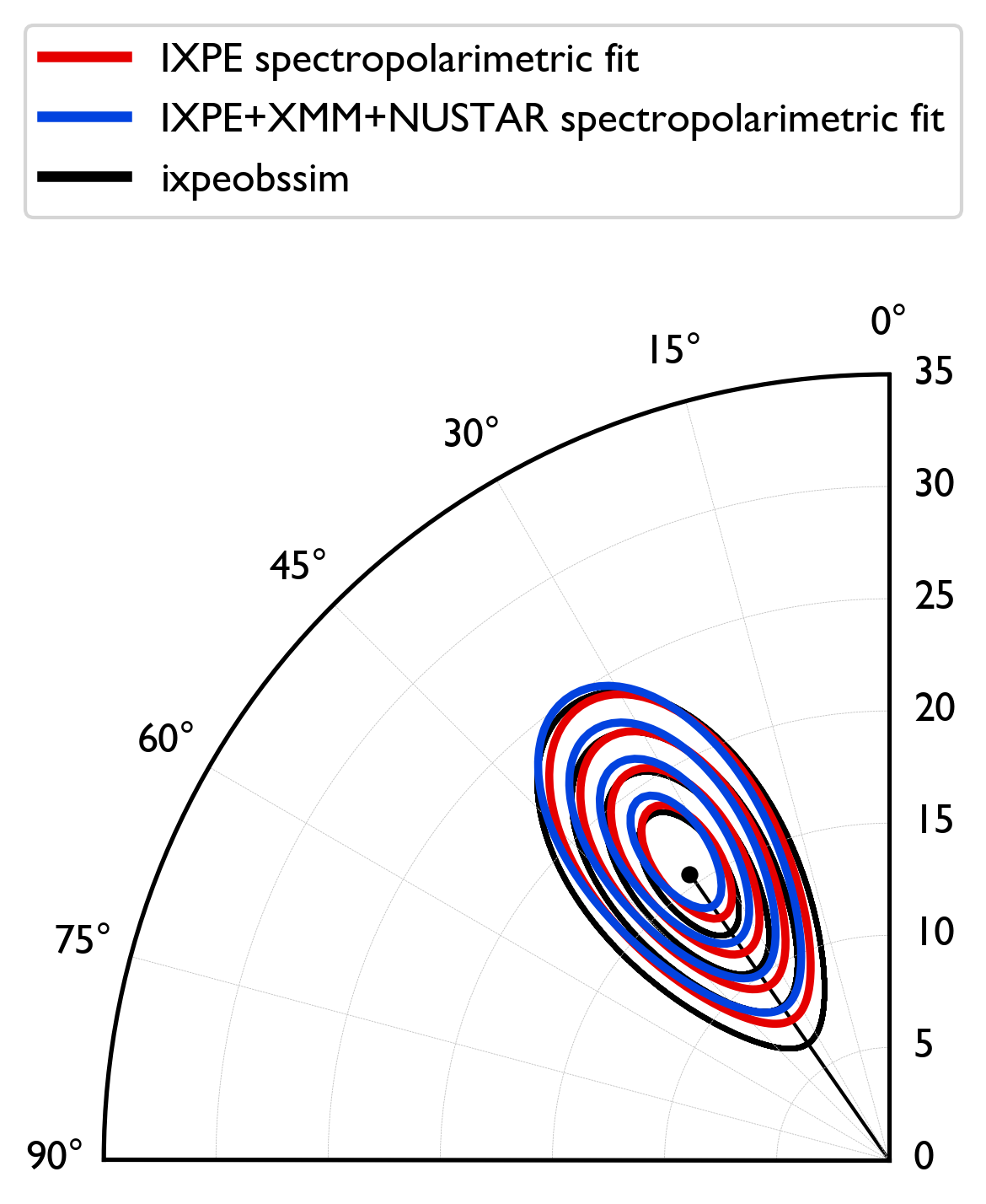} }
 \caption{Polar plot of the $\Pi_{\rm x}$-$\psi_{\rm x}$ plane. We show the confidence contours of X-ray polarization derived with ixpeobssim (black), a spectro-polarimetric fit of the IXPE data (red) and
 a spectro-polarimetric fit of the {\it IXPE}, {\it XMM} EPIC-pn, and {\it NuSTAR} data (blue). For each case, the contours at 68.3\%, 95.4\%, 99.7\%, and 99.994\% confidence level are shown.}
\label{allcontour.fig}
\end{figure}

Next, to determine the polarization properties from the polconst model, we added to the fit the IXPE Stokes $Q$ and $U$ spectra. This fitting exercise returned an estimation of the polarization parameters of $\Pi_{\rm x}=16\pm2\%$ and $\psi_{\rm x}=35\pm3\degr$, which is consistent with the ixpeobssim result and with the fit using only {\it IXPE} data. As a final check, we tested whether allowing the spectral shape to differ between {\it NuSTAR}, {\it XMM} Epic-pn, and {\it IXPE} could further improve the fit and thereby affect the values of $\Pi_{\rm x}$ and $\psi_{\rm x}$. Decoupling the $\alpha$ and $\beta$ parameters among the three instruments resulted in a statistically significant improvement of the fit ($\Delta \chi^2=-72$ for 4 more d.o.f., corresponding to a decrease in the Bayesian information criterion $\Delta \rm BIC$=-43). This discrepancy in spectral shape between the three telescopes is likely due to non-simultaneity of the observations as the source varies (see \S\S\ref{var.sec}). We note however, that this fitting exercise returned the same result, within the uncertainty, for the polarization properties: $\Pi_{\rm x}=16\pm2\%$, $\psi_{\rm x}=34\pm4\degr$. This indicates that the polarization determination is robust against
differences in the details of the spectral shape modeling. Our final spectro-polarimetric fit, including {\it XMM} EPIC-pn, {\it NuSTAR}, and {\it IXPE}, is displayed in Fig. \ref{specpol.fig}. To highlight the significance of the polarization detection and the agreement among the determinations via different methods that use different models for the spectral curvature in Fig. \ref{allcontour.fig}, we compared the two-dimensional confidence contours for the X-ray polarization obtained with ixpeobssim and with our spectropolarimetric fits.

Finally, for the MLS method, we fit the {\it IXPE} data with a power-law model with Galactic photo-electric absorption (see above), while holding the intrinsic values of $Q$ and $U$ constant. In this way, we derived a photon index $\Gamma=2.97\pm0.01$ and polarization properties ($\Pi_{\rm x}=16\pm2\%$ and $\psi_{\rm x}=34\pm2\degr$) consistent with the values from the previously described methods. In addition, using a model-independent maximum-likelihood calculation, we obtained $\Pi_{\rm x}=16\pm2\%$ and $\psi_{\rm x}=34\pm2\degr$. This is a further, independent confirmation of our polarization measurement of Mrk 421.

%
%
%

%
\subsection{Polarization variability}
\label{var.sec}
\begin{figure}
\resizebox{\hsize}{!}{\includegraphics[scale=0.05]{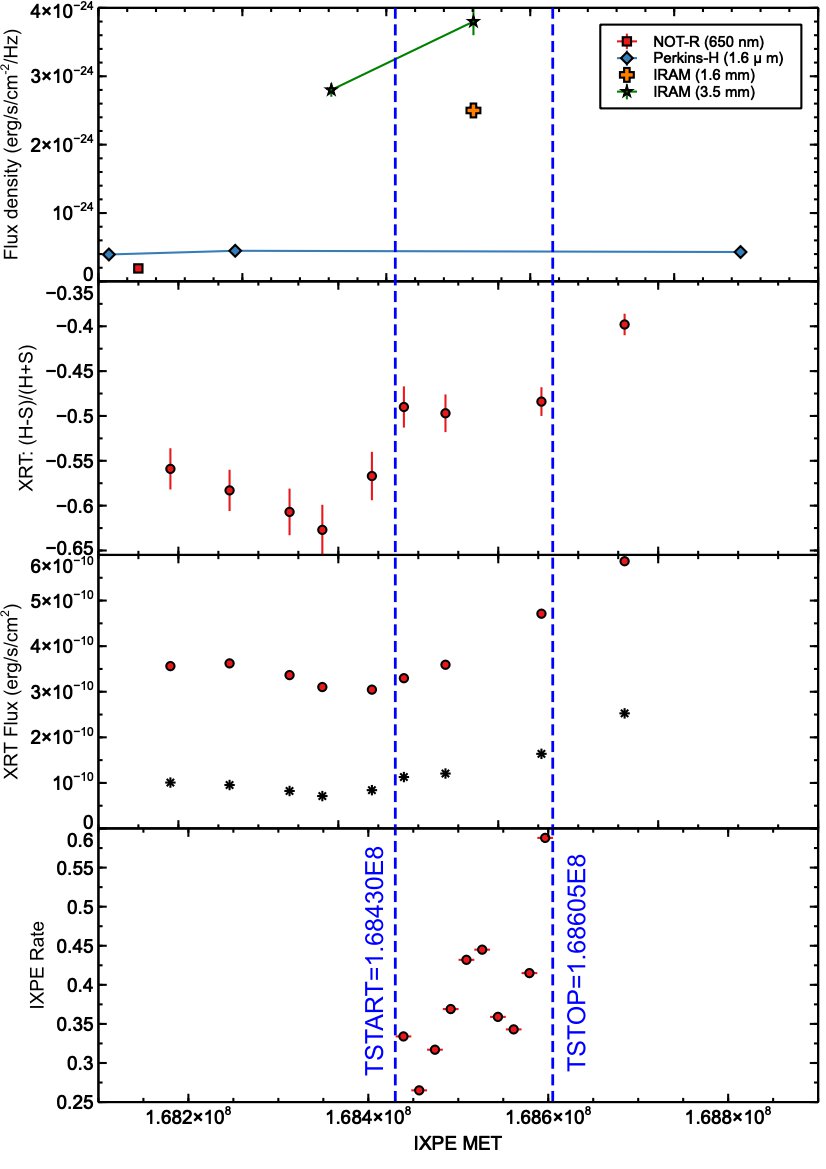} }
 \caption{From the top to the bottom: light-curves of the optical, infrared and millimeter flux densities, light-curve of {\it Swift} XRT hardness ratio,
 light-curve of the soft (S band: 0.3-2.0 keV) and hard (H band: 2.0-10.0 keV) {\it Swift} XRT fluxes, and light-curve of the {\it IXPE} count rates. In the first panel, the legend displays the symbols for the different instruments and wavelenghts. Multiple data points from the same instrument are connected by a line to guide the eye. In the third panel, soft and hard fluxes are coded as red circles and blue asteriks. The dashed vertical lines indicate the beginning and the end of the {\it IXPE} observation.}
\label{lc.fig}
\end{figure}

To obtain additional information about the physical conditions in the jet of Mrk~421, we have searched for variability of the X-ray polarization properties as a function of time within the total duration of the {\it IXPE} pointing. We first exploited the {\it Swift} XRT monitoring of the source (9 separate pointings, each for $\sim1$ ks) to sample the spectral evolution of the source before, during, and after the {\it IXPE} pointing. As a comparison, we created a light curve of the {\it IXPE} count rate with time bins of 10 ks (Fig. \ref{lc.fig}). The flux of the source increased during the {\it IXPE} pointing in both the soft (0.3-2 keV) and hard (2-10 keV) XRT bands, while maintaining the same spectral slope, as witnessed by the hardness ratio light curve. In the {\it IXPE} data, we find an increase of the count rate by a factor of 2.2.

To investigate whether there was also a variation in the polarization properties on different time-scales within the {\it IXPE} pointing, we proceeded as follows. Using time bins in the range 5-50 ks, we created light curves of the normalized Stokes parameters. For each case, we fit the light curve with a constant model and recorded the probability of the null hypothesis of obtaining a value of $\chi^2$ at least as large as that of the constant model. For all of the time bins tested, we found that this probability is always above 1\%, which indicates that the Stokes parameters are consistent with no variation during the {\it IXPE} pointing.  We therefore conclude that, despite the increase in flux during the observation, the polarization properties did not vary substantially. As a useful quantitative constraint, we computed the uncertainty in $\Pi_{\rm x}$ and $\psi_{\rm x}$ allowed by the statistic in each time bin. This is an indication of the amount of variation that can be excluded, on different time-scales, by the present analysis. On a time-scale of 50 ks, similar to 1-day time-scales sometimes used in studies of optical polarization variability \citep[e.g.,][]{blinov15}, variations in $\Pi_{\rm x}$ above $\sim\pm4\%$ for a constant value of $\psi_{\rm x}$, or variations in $\psi_{\rm x}$ above $\sim\pm7\degr$ for constant $\Pi_{\rm x}$, are excluded by the present analysis.

\begin{figure}
\resizebox{\hsize}{!}{\includegraphics[scale=0.05]{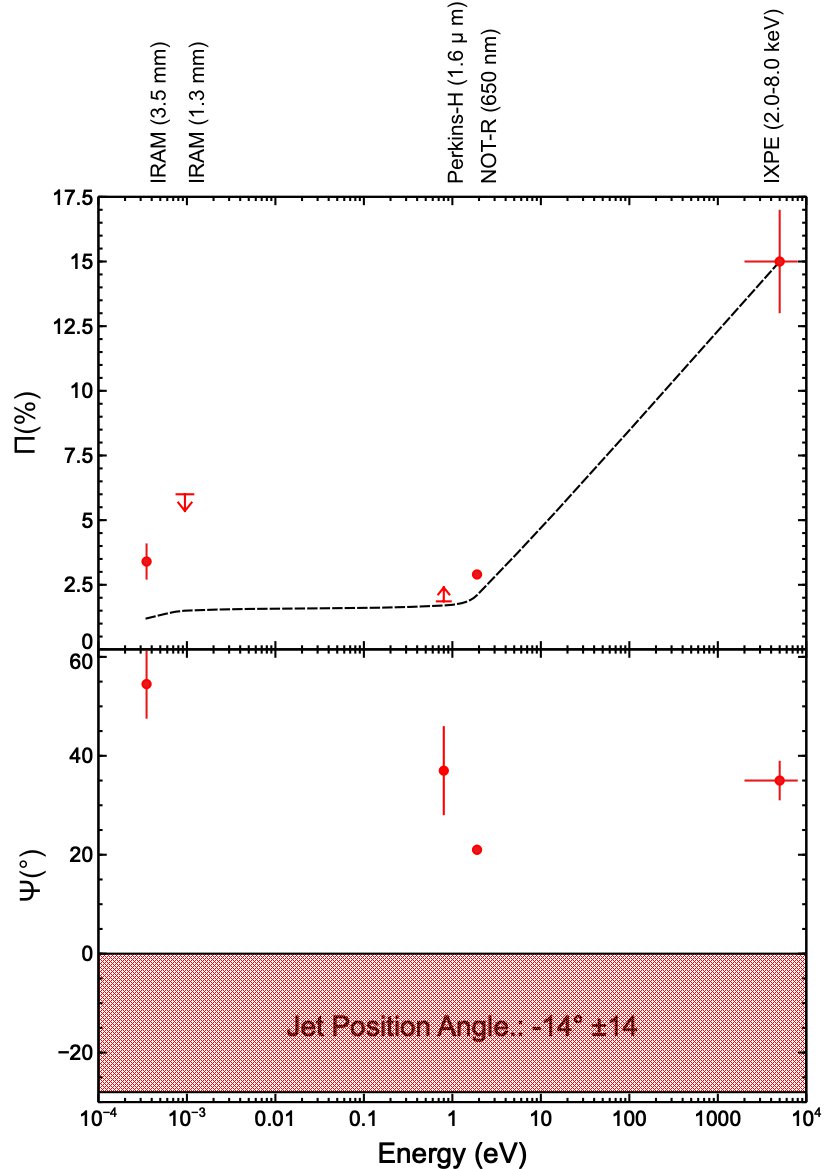} }
\caption{Multi-wavelength polarization degree (top panel) and angle (low panel) of Mrk~421 as a function of energy. The red circles represent detections, while the arrows indicate upper or lower limits. The instruments that provided each measurement are labeled on top. The dashed line represents theoretical estimations assuming the formalism of \citet{hughes1991} for a turbulent jet. In the second panel, the horizontal shaded area indicates the time-averaged jet position angle and its uncertainty, measured with the Very Long Baseline Array at 43 GHz \citep{weaver2022}.}
\label{allpol.fig}
\end{figure}
%

\section{Discussion}\label{sec:disc_conc}
In optically-thin synchrotron sources, such as Mrk~421 at IR to X-ray wavelengths, the degree of polarization $\Pi$ is a measure of the level of disorder of the magnetic field, while $\psi$ is perpendicular to the mean magnetic field direction as projected onto the sky plane. The dependence of the polarization on wavelength relates to the mechanism that accelerates the particles in situ. 
In a highly magnetized jet, conversion of magnetic into kinetic energy is a viable mechanism to accelerate the particles. If the emission is modulated by the development of a large-scale kink instability, relatively high (e.g., $\sim$20\%) X-ray and optical polarization is predicted, with both $\Pi$ and $\psi$ smoothly modulated with time (e.g., \citealt{bodo21}). On the other hand, when single current sheets are responsible for most of the observed emission, large and rapid variability is expected, leading to substantial dilution of the polarization when integrated over long exposures. In this case, one expects to measure a lower degree of polarization in the X-rays than in the optical band \citep{zhang2021}.

In contrast, in the case of weakly magnetized jets, a shock is a more likely acceleration mechanism.
According to several models \citep{marscher1985,2018MNRAS.480.2872T,angelakis16}, if particles are accelerated at a shock front and advected downstream, the emission at the shortest (X-ray) wavelength is produced by energetic particles that are present only very close to the shock front. Longer-wavelength emission extends over a larger volume that contains the lower-energy particles. 
If the magnetic field direction changes, or level of disorder increases, over the larger volume downstream of the acceleration site, $\Pi_{\rm x}$ at shorter wavelengths is expected to be higher than the degree of polarization at longer wavelengths. If the plasma in the jet flow is highly turbulent before crossing the shock, the field will be partially ordered parallel to the shock front, while the flux, $\Pi$, and $\psi$ should be time-variable (although the average value of $\psi$ should remain parallel to the jet axis), more strongly so at X-ray than at longer wavelengths \citep{marscher14}. Under these conditions, the erratic variability of both degree and angle of polarization can potentially lead the X-ray polarization vectors to average to low values in day-long observations \citep{digesu2022}. 
If the flow only becomes turbulent after crossing the shock, the time variability can be weak, but $\Pi$ should decrease toward longer wavelengths.  If we assume that synchrotron losses dominate the radiative energy losses, the timescale of the losses is inversely proportional to the Doppler factor, $\delta$, frequency, $\nu$, and strength of the magnetic field, $B$, as $\tau_{\rm loss}\propto [(1+z)/(\delta\nu B^3)]^{1/2}$ \citep{marscher1985}. If $\delta$ and $B$ are the same across the X-ray/optical/IR regions (and possibly less in the site of millimeter-wave emission), the size of the regions behind the shock with respect to the size of the X-ray emission region can be determined as $R_{\nu}=R_{X}(\nu_{X}/\nu)^{1/2}$. If we consider the magnetic field be turbulent, according to the formalism suggested by \citet{hughes1991}, the IXPE measurement implies that the X-ray emission should consist of $\sim$30 cells with different magnetic field directions. Therefore, within this interpretation, we can estimate the sizes (in terms of number of cells) and polarization of 
the optical, IR, 1.3mm, and 3mm regions. We find predicted values of $\Pi$ equal to 2.1\%, 1.7\%, 1.5\%, and 1.2\% in R band, H band and at 1.3mm and 3mm, respectively.

In Figure \ref{allpol.fig} we summarize the multi-wavelength polarization properties that we have measured for Mrk~421 during an average state in 2022 May. The degree of polarization in the X-ray band is higher by a factor of $\sim$5 than in the optical and by at least a factor $\sim$3 than at IR and millimeter wavelengths.  
This is in qualitative agreement with the energy-stratified shock model described above, although the theoretically predicted values of $\Pi$ are somewhat lower than the measured values if we use the degree of X-ray polarization as a reference. In addition, shock acceleration is favored by the observed temporal stability of the polarization properties (\S \ref{var.sec}). This is a natural prediction of, e.g., a shock model with a self-generated magnetic field \citep{tavecchio21}, while it is in tension with magnetic reconnection models, which predict smooth \citep{bodo21}, or even rapid \citep{zhang2018}, polarization variability, including polarization angle swings.

One can see in Figure \ref{allpol.fig} that there is a discrepancy between the multi-wavelength $\psi$ values and the time-averaged\footnote{See \url{http://www.bu.edu/blazars/VLBA_GLAST/1101.html} for the multi-epoch variability of the jet position angle.} jet position angle measured at 43 GHz with the Very Long Baseline Array ($-14\degr \pm 14\degr$, \citealt{weaver2022}).  In contrast, although there are exceptions, the optical polarization position angle of BL Lac objects is frequently found to align (within $\pm20\degr$) with that of the core of the jet \citep{marscher21}. This has been ascribed to an ordered component of the magnetic field (e.g. a helical magnetic field compressed by a shock) aligned perpendicular to the jet axis. However, the jet of Mrk~421 has a wide opening angle of $\sim 60$\degr, owing mainly to projection effects caused by a small angle to the line of sight \citep{weaver2022}. It is possible that the jet bends by tens of degrees on the smallest scales, such that there is a closer correspondence between $\Psi$ and the jet direction where the X-ray and optical emission occurs. Future very long baseline interferometry at 1 mm wavelength can determine whether this is the case. Misalignments between the polarization angle and the jet direction, possibly induced by local disturbances of the magnetic field, have been reported for the variable radio knots of Mrk~421 \citep{piner2004}.
\section{Conclusions}

The primary findings of the first {\it IXPE} observations of X-ray polarization from Mrk~421 in
an average flux states are that (1) the degree of X-ray polarization was $15\pm2\%$, several times higher than at optical/infared/millimeter wavelengths; (2) the polarization properties were steady, while the X-ray flux rose by a factor of 2.2, during the $\sim 2$-day {\it IXPE} pointing; and (3) the X-ray electric-vector position angle was $+35\pm4\degr$, close to the optical value of $+21\pm1\degr$, but different from the jet direction of $-14\pm14\degr$ measured at 43 GHz. 

The finding of high X-ray polarization compared to longer wavelengths implies that the particles in the jet of Mrk~421 are energy-stratified. The higher-energy particles responsible for the emission at higher (X-ray) frequencies are located closer to the acceleration site, since their energy decays more rapidly than is the case for lower-energy particles. As the particles stream away from the acceleration site, their energies decrease, so there is a negative gradient of maximum frequency of synchrotron radiation with distance from the acceleration region. If the particles encounter a more disordered magnetic field as they advect downstream, the degree of polarization then decreases toward longer wavelengths. The lack of polarization variation as the X-ray flux increases suggests that particles are injected in a region with stable magnetic field properties. Partially ordered magnetic fields can be produced close to the shock because of plasma processes \citep{2018MNRAS.480.2872T} or compression \citep{hughes1985}. In this case, $\psi$ should align with the jet axis, as found in Mrk 501 (Liodakis et al. 2022, submitted).
The increase in the level of disorder of the field with distance from the shock can be caused by turbulence. However, our finding of little or no variation of the X-ray polarization properties on a timescale of 2 days,
in contrast with the finding of monitoring of the optical polarization variability \citep{marscher21}, suggests that the IXPE observation
occurred during a time when the turbulence happened to be strong but stable.

Given the above expectations of theoretical models, our findings imply that shock acceleration is the most likely particle acceleration mechanism in the jet of Mrk~421 during the observation reported here. The \virg{harder-when-brighter} spectral behavior that is commonly displayed by HSPs in the X-ray band, and that we also observed in our Swift monitoring of Mrk~421 (Fig. \ref{lc.fig}), is widely interpreted as due to the injection of fresh high-energy electrons by a shock \citep{kirk1998}.

The discrepancy that we have found between the radio jet position angle and the millimeter/optical/infrared/X-ray polarization angle suggests that either the radio emission region is detached from the X-ray emission region, as is common in one-zone models where the X-ray emitting region is opaque at radio frequencies \citep[e.g.,][]{ghisellini09}, or that the jet bends between the sites of high- and low-frequency emission \citep{Marscher2008}.
Using the formalism of \citet{hughes1991} for a turbulent jet, we have estimated the optical/infrared/millimeter $\Pi$ values 
that are implied by the X-ray measurement.
These are somewhat lower than the observed values (Fig. \ref{allpol.fig}), which suggests that the magnetic field behind the shock is not completely turbulent, but instead includes an ordered component.

Our study is part of a fundamental early attempt to explore the polarization properties of blazars over a wide range of wavelengths, now including X-rays. An important aspect of this is to determine whether, and in what manner, the polarization varies. However, the dearth of exactly simultaneous sampling at X-ray and optical wavelengths in the present study, and the limited sensitivity of the {\it IXPE} data to variations on short time-scales, prevent us from drawing firm general conclusions from this single realization.
Multi-wavelength polarization monitoring of blazars on different time-scales is needed to probe more thoroughly the particle distribution and magnetic field dynamics in astrophysical jets, for example by searching for correlated variability patterns and time delays between the polarization light curves in different bands. In the future, the X-ray polarization information provided by {\it IXPE} will play a key role in these types of studies by adding to the existing probes valuable information on the location and physical conditions experienced by freshly accelerated high-energy particles.

\begin{acknowledgments}
The Imaging X ray Polarimetry Explorer (IXPE) is a joint US and Italian mission.  The US contribution is supported by the National Aeronautics and Space Administration (NASA) and led and managed by its Marshall Space Flight Center (MSFC), with industry partner Ball Aerospace (contract NNM15AA18C).  The Italian contribution is supported by the Italian Space Agency (Agenzia Spaziale Italiana, ASI) through contract ASI-OHBI-2017-12-I.0, agreements ASI-INAF-2017-12-H0 and ASI-INFN-2017.13-H0, and its Space Science Data Center (SSDC), and by the Istituto Nazionale di Astrofisica (INAF) and the Istituto Nazionale di Fisica Nucleare (INFN) in Italy.  This research used data products provided by the IXPE Team (MSFC, SSDC, INAF, and INFN) and distributed with additional software tools by the High-Energy Astrophysics Science Archive Research Center (HEASARC), at NASA Goddard Space Flight Center (GSFC).
  The IAA-CSIC group acknowledges financial support from the Spanish \virg{Ministerio de Ciencia e Innovaci\'on} (MCINN) through the \virg{Center of Excellence Severo Ochoa} award for the Instituto de Astrof\'isica de Andaluc\'ia-CSIC (SEV-2017-0709) and through grants AYA2016-80889-P and PID2019-107847RB-C44. The POLAMI observations were carried out at the IRAM 30m Telescope. IRAM is supported by INSU/CNRS (France), MPG (Germany) and IGN (Spain).
  Some of the data reported here are based on observations made with the Nordic Optical Telescope, owned in collaboration by the University of Turku and Aarhus University, and operated jointly by Aarhus University, the University of Turku and the University of Oslo, representing Denmark, Finland and Norway, the University of Iceland and Stockholm University at the Observatorio del Roque de los Muchachos, La Palma, Spain, of the Instituto de Astrofisica de Canarias. E. L. was supported by Academy of Finland projects 317636 and 320045. The data presented here were obtained [in part] with ALFOSC, which is provided by the Instituto de Astrofisica de Andalucia (IAA) under a joint agreement with the University of Copenhagen and NOT. Part of the French contributions is supported by the Scientific Research National Center (CNRS) and the French spatial agency (CNES). The research at Boston University was supported in part by National Science Foundation grant AST-2108622, NASA Fermi Guest Investigator grant 80NSSC21K1917, and NASA Swift Guest Investigator grant 80NSSC22K0537. This research was conducted in part using the Mimir instrument, jointly developed at Boston University and Lowell Observatory and supported by NASA, NSF, and the W.M. Keck Foundation. We thank D. Clemens for guidance in the analysis of the Mimir data.
\end{acknowledgments}

\software{Xspec} version 12.12.1
\facilities{{\it IXPE}, Nordic Optical Telescope, {\it NuSTAR} {\it Swift}, {\it XMM-Newton}, Perkins Telescope Observatory}

\appendix


\section{IXPE and multi-wavelength data analysis} 
\label{sec:IXPE_data}

\subsection{{\it IXPE} data}
{\it IXPE} data are processed using a pipeline\footnote{\url{https://heasarc.gsfc.nasa.gov/docs/ixpe/analysis/IXPE-SOC-DOC-009-UserGuide-Software.pdf}} that estimates the photo-electron emission direction, correcting for charging effects, detector temperature, gas electron multiplier (GEM) gain non-uniformity, and polarization artifacts induced by spurious instrumental modulation \citep{rankin2022}. The level-2 event files (one for each of the three DUs) produced by the pipeline contain all of the information typical of an imaging X-ray astronomy mission (e.g., photon arrival time, detector coordinates, and energy), with the addition of the polarization information in the form of event-by-event Stokes parameters. Before proceeding with the science analysis, we corrected the data for small time-dependent changes to the gain correction obtained from data taken with the on-board calibration sources \citep{Ferrazzoli2020JATIS} close to the actual time of observation.\\

We performed the scientific analysis of the {\it IXPE} data using the ixpeobssim software \citep{pesce2019, baldini2022}. At the angular resolution of {\it IXPE} ($\sim30''$), blazars like Mrk~421 are point-like sources. 
The selection of {\it IXPE} events was performed with the xpselect tool, which applies to the photon lists a user-defined selection on sky position, time, phase, and energy. In the case of Mrk~421, we extracted the source events using circular regions 60\arcsec in radius, while we used an annulus centered on the source, with an in inner and outer radius of 2\arcmin and 5\arcmin, respectively, to estimate the background. We found that the background accounts for less than 2\% of the total counts in the source region, hence background subtraction is not critical for the polarization measurement.
The \citet{Kislat2015} method for estimating the polarization of a user-defined set of events is implemented in the PCUBE algorithm.
We created the Stokes-parameter spectra of the source using the PHA1, PHAU, and PHAQ alghoritms, which map the $I$, $Q$, and $U$ Stokes parameters of the photons into OGIP-compliant PHA spectral files (3 Stokes-parameter spectra per 3 detector units). We prepared the spectra for the spectro-polarimetric fit by binning the $I$ spectrum, requiring that a minimum of 30 counts is reached in each energy bin. This is needed for correctly using the \chis statistics in the fit. Hence, as a last step, we applied the same binning to the $Q$ and $U$ spectra. 

\subsection{Spectroscopic X-ray data}
In order to catch any possible X-ray flux variations of the source contemporaneous with the {\it IXPE} pointing, and to accurately constrain the slope of the X-ray spectrum, we observed Mrk~421 in the 0.3-10 keV energy band with the {\it Swift} XRT (9 pointings, each with exposure time $\sim1$ ks, of which one was simultaneous with {\it IXPE}), and with {\it NuSTAR} in the 3-70 keV energy range (see Table \ref{obs.tab}). In addition, an {\it XMM-Newton} (5 ks) observation was carried out as a part of our program on 2022 May 3, one day before the {\it IXPE} pointing.\\
The {\it Swift} XRT observations were performed in Windowed Timing (WT) mode,
and the data were processed with the X-Ray Telescope Data Analysis Software
(XRTDAS, v. 3.6.1). In the analysis, we used the latest calibration files
available in the {\it Swift} XRT CALDB (version 20210915). The X-ray source spectrum was extracted from the cleaned event file using a circular region with a radius
of 47\arcsec. The background was extracted using a circular region with the same
radius from a blank-sky WT observation available in the {\it Swift} archive. As a final step, we binned the 0.3-10 keV channels to achieve at least 25 counts in each energy bin.\\
We calibrated and cleaned the {\it NuSTAR} data using the {\it NuSTAR} Data Analysis Software (NuSTARDAS, v. 2.1.1) and the latest calibration data files in the {\it NuSTAR} CALDB (version 20220301). 
The net exposure time after cleaning the dataset was 25 ks.
For both the FPMA and FBMB detectors, we created level 2 cleaned event files with the
nupipeline task, while high-level scientific data products were extracted using the nuproducts tool. We extracted the source spectrum from a circular region with a radius of 70\arcsec, while we used an annular region to estimate the background. Finally, we binned the spectra to achieve at least 50 counts in each energy bin.\\

The {\it XMM-Newton} observation of Mrk~421 was obtained in timing mode, which limits photon pile-up. We performed the data reduction using the XMM-Newton Science Analysis Software (SAS), version 20, and the latest calibration products. Starting from the raw observation data files (ODFs), we created the calibrated EPIC-pn event list using the SAS task epchain. To determine good time intervals (GTIs) free from background flares, we cut the time intervals where the light-curve of the background in the 10-13 keV band was above a fixed treshold of 0.4 counts/s (SAS task tabgtigen). In the clean event files, we selected the source and the background using boxed regions (with a width of 27 pixels) that maximize the signal-to-noise ratio. We created the source and background spectra using only single and double events, and we created the spectral response matrices using the SAS task rmfgen and arfgen. Finally, we prepared the spectrum for analysis by binning the energy channels to achieve at least 30 counts in each spectral bin. The model fitting to the spectrum was carried out with Xspec version 12.12.1, as described in the main text. To allow simultaneous spectro-polarimetric fitting with EPIC-pn and {\it NuSTAR}, we added to keyword "XFLT0001 Stokes:0" to the header of the non-{\it IXPE} pha files.
This allowed Xspec to identify the spectra as I Stokes spectra.
\subsection{Optical and infrared data}
Optical polarization observations were performed using the Alhambra Faint Object Spectrograph and Camera (ALFOSC) mounted at the 2.5 m Nordic Optical Telescope (NOT) on the night of 2022 May 4 (MJD 59703) in the R band ($\lambda$=6500 \AA/$E$=1.9 eV, \fwhm=1300 \AA /0.4 eV). The data were obtained in the standard polarimetric mode and analyzed using a 2.5$''$ aperture radius and standard photometric procedures included in the pipeline developed at Tuorla Observatory \citep{Hovatta2016,Nilsson2018}. Unpolarized and polarized standard stars were also used for calibration. The R magnitude of the source was 12.98$\pm0.02$, which corresponds to $I=19.8\pm0.3$ mJy, while $\Pi_{\rm O,obs}=2.42\pm0.06\%$ and $\psi_{\rm O}=21\pm1$. For the 2.5$''$ aperture used in the analysis, the host galaxy contributes $I_{host}=3.2\pm0.4$ mJy to the total emission \citep{Nilsson2007}. We correct for the host galaxy contamination of the polarization degree by subtracting the host-galaxy flux density as $\Pi_\mathrm{intr}= \Pi_\mathrm{obs}\times{I}/(I-I_\mathrm{host})$ \citep{Hovatta2016}. The intrinsic polarization is therefore $\Pi_{\mathrm O,int}=2.9\pm0.1\%$.\\
Infrared observations in the H band were obtained at the Boston University 1.8 m Perkins telescope located in Flagstaff, AZ (Perkins Telescope Observatory, PTO, BU) using the IR camera MIMIR\footnote{\url{https://people.bu.edu/clemens/mimir/index.html}} on 2022 May 5, 6, and 9. Each observation consisted of 6 dithering exposures of 5 s each at 16 positions of a half-wave plate, rotated in steps of 22.5$\degr$ from 0 to 360 $\degr$. The camera and data reduction procedures are described in detail by \citet{clemens2012}
The H-band degree of polarization changed from 1\% to 2.2\%, with average 1$\sigma$ uncertainty of 0.5\%. However, the $\Pi$ measurements in H band are lower limits, since the level of contamination from the host galaxy is unknown. The position angle of polarization in H band agrees within the uncertainty with that of the IXPE observation.

%

\subsection{Millimeter-wave data}

Radio polarization at 3.5 mm (86.24 GHz) and
1.3 mm (230 GHz) was measured with the 30 m Telescope of the Institut
de Radioastronomie Millim\'etrique (IRAM), located at the Pico Veleta
Observatory (Sierra Nevada, Granada, Spain), on 2022 May 5 and 7 (MJD
59704 and 59706), within the Polarimetric Monitoring of AGN at Millimeter
Wavelengths (POLAMI) program \citep{polami_i,polami_iii}. 
Under the POLAMI\footnote{\url{http://polami.iaa.es/}}
observing setup, the four Stokes parameters ($I$, $Q$, $U$, and $V$) 
are recorded simultaneously using the XPOL procedure \citep{xpol}. 
We reduced, calibrated, and managed the data following
the POLAMI procedure described in detail in \citet{polami_i}. 
In the two measurements at 3.5 mm, the source flux density increased
from $0.278\pm0.014$ Jy to $0.384\pm0.019$ Jy, while at the same time
the polarization properties remained relatively stable (see Table \ref{obs.tab}). Only one measurement was taken at 1.3 mm, on 2022 May 7, providing an upper limit
at 95\% confidence level of 6\% for the degree of polarization.


\end{document}